\documentclass{ws-ijmpa}
\usepackage{amsmath}
\def\epsi{\varepsilon}
\def\khat{\widehat K}
\begin{document}

%%%%%%%%%%%%%%%%%%%%%%%%%%% To switch off trimmarks %%%%%%%%%%%%%%%%%%%
\def\nocropmarks{\vskip5pt\phantom{cropmarks}}
\let\trimmarks\nocropmarks  
\markboth{Shunzo Kumano}
{A model prediction for polarized antiquark flavor asymmetry}
%%%%%%%%%%%%%%%% Publisher's Area please ignore %%%%%%%%%%%%%%%
\catchline{}{}{}
%%%%%%%%%%%%%%%%%%%%%%%%%%%%%%%%%%%%%%%%%%%%%%%%%

%%%%%%%%%%%%%%%%%%%%%%%%%%%%%%%%%%%%%%%%%%%%%%%%%%%%%%%%%%%%%%%%%%%%%%%%%%%%%%%%
%%%%%%%%%%%%%%%%%%%%%%%%%%%%%%%%%%%%%%%%%%%%%%%%%%%%%%%%%%%%%%%%%%%%%%%%%%%%%%%%
\pagestyle{empty}
\begin{flushleft}
\Large{SAGA-HE-183-01    \hfill December 8, 2001}
\end{flushleft}
\vspace{1.5cm}
 
\begin{center}
\LARGE{\bf A model prediction for polarized antiquark flavor asymmetry} \\
\vspace{1.2cm}
\Large{ S. Kumano $^*$}  \\
\vspace{0.4cm}
{Department of Physics \\
 Saga University \\
 Saga, 840-8502, Japan} \\
\vspace{1.8cm}
\Large{Talk given at the 3rd Circum-Pan-Pacific Symposium on \\
      ``High Energy Spin Physics"} 

\vspace{0.3cm}
{Beijing, China, Oct. 8-13, 2001} \\
{(talk on Oct. 11, 2001) }  \\
\end{center}
\vspace{1.6cm}
\noindent{\rule{6.0cm}{0.1mm}} \\
\vspace{-0.3cm}
\normalsize

\noindent
{* Email: kumanos@cc.saga-u.ac.jp;
   \ \, URL: http://hs.phys.saga-u.ac.jp.}  \\

\vspace{+0.1cm}
\hfill {\large to be published in proceedings}
\vfill\eject
\pagestyle{plain}

%%%%%%%%%%%%%%%%%%%%%%%%%%%%%%%%%%%%%%%%%%%%%%%%%%%%%%%%%%%%%%%%%%%%%%%%%%%%%%%%
%%%%%%%%%%%%%%%%%%%%%%%%%%%%%%%%%%%%%%%%%%%%%%%%%%%%%%%%%%%%%%%%%%%%%%%%%%%%%%%%

\setcounter{page}{1}

\title{A MODEL PREDICTION \\ FOR POLARIZED ANTIQUARK FLAVOR ASYMMETRY}

\author{S. KUMANO}
\address{Department of Physics, Saga University,
              Saga, 840-8502, Japan \\
              Email: kumanos@cc.saga-u.ac.jp \\
              URL: http://hs.phys.saga-u.ac.jp}
\maketitle

%\pub{Received (received date)}{Revised (revised date)}

\begin{abstract}
Polarized flavor asymmetry $\Delta \bar u/\Delta \bar d$ is investigated
in a meson-cloud model. A polarized nucleon splits into a $\rho$ meson
and a baryon, then the polarized $\rho$ meson interacts with the virtual
photon. Because of the difference between the longitudinally polarized
distributions $\Delta \bar u$ and $\Delta \bar d$ in $\rho$, 
the polarized flavor asymmetry is produced in the nucleon. 
In addition, we show that the $g_2$ part of $\rho$ contributes
to the asymmetry especially at medium $x$ with small $Q^2$. 
\end{abstract}

%%%%%%%%%%%%%%%%%%%%%%%%%%%%%%%%%%%%%%%%%%%%%%%%%%%%%%%%%%%%%%%%%%%%%%%%%%%%%%
\section{Introduction}\label{intro}

Spin structure of the nucleon has been investigated extensively
for the last ten years, and now we have a rough idea on the internal
spin structure. The experimental information comes mainly from
inclusive lepton scattering experiments.
Although there are polarized semi-inclusive data, they are
not accurate enough to pose a strong constraint, for example, on
the polarized flavor asymmetry $\Delta \bar u/\Delta \bar d$.
However, it will be clarified experimentally in the near future
by $W$ production experiments at RHIC and also semi-inclusive measurements
by the COMPASS collaboration. 

Most theoretical papers on the unpolarized asymmetry $\bar u/\bar d$ are
written after the NMC discovery on Gottfried-sum-rule violation.
Therefore, the unmeasured $\Delta \bar u/\Delta \bar d$ is an appropriate
quantity for testing various theoretical models. In this sense, it is
desirable to present model predictions before the experimental data will
be taken. There are already some model predictions by Pauli-exclusion,
chiral-soliton, and meson-cloud models. The meson models are successful
in explaining the unpolarized asymmetry,\cite{skpr}
so that we try to investigate the details of the model
in the polarized asymmetry.\cite{fs}$^-$\cite{cw}
The following discussions are based on Ref. 4.

%%%%%%%%%%%%%%%%%%%%%%%%%%%%%%%%%%%%%%%%%%%%%%%%%%%%%%%%%%%%%%%%%%%%%%%%%%%%%%
\section{$\rho$ meson contributions}\label{summary}

We explain the outline of the formalism for calculating $\rho$ meson
contributions to the flavor asymmetric distribution
$\Delta \bar u - \Delta \bar d$. 
The polarized nucleon splits into a $\rho$ meson and a baryon, then
the virtual photon interacts with the polarized $\rho$ meson.
The $\rho$ meson is a spin-1 hadron, and $\rho^+$ or $\rho^-$
has difference between $\bar u$ and $\bar d$. Therefore, it affects
the polarized flavor asymmetry $\Delta \bar u - \Delta \bar d$
in the nucleon.

The contribution to the nucleon tensor $W_{\mu \nu}$ from the splitting
process into a vector meson $V$ and a baryon $B$ is expressed as\cite{km}
\begin{equation}
W_{\mu\nu} (p_N, s_N, q)  =  \int \frac{d^3 p_B}{(2\pi)^3}
            \, \frac{2 m_V m_B}{E_B} 
   \sum_{\lambda_V,\lambda_B} | J_{VNB} |^2
            \, W_{\mu\nu}^{(V)} (k, s_V, q)  .
\label{eqn:w-meson}
\end{equation}
Here, $m_V$ and $m_B$ are the meson and baryon masses, 
$p_N$, $p_B$, $k$, and $q$ are the nucleon, baryon, meson, and
virtual photon momenta, $s_N$ and $s_V$ are the nucleon and meson spins,
$J_{VNB}$ is the $VNB$ vertex multiplied by the meson propagator,
and $W_{\mu\nu}^{(V)}$ is the meson tensor.
Polarized structure functions $g_1$ and $g_2$ are defined in
the antisymmetric part of the nucleon tensor:
\begin{equation}
W^A_{\mu \nu} (p_N, s_N, q) = 
       i \, \varepsilon_{\mu\nu\rho\sigma} \, q^{\, \rho}
\bigg [ \,  s_N^{\, \sigma} \, \frac{g_1}{p_N \cdot q} 
        + ( p_N \cdot q \, s_N^{\, \sigma} - s_N \cdot q \, p_N^{\, \sigma} ) 
          \, \frac {g_2}{(p_N \cdot q)^2} \,
\bigg ] .
\label{eqn:g1g2}
\end{equation}
In order to separate $g_1$ from $g_2$, the projection operator
\begin{equation}
P^{\mu \nu} = - \frac{m_N^2}{2 \, p_N \cdot q} 
         \, i \, \varepsilon^{\mu\nu\alpha\beta} \, q_\alpha \, s_{N \, \beta} ,
\end{equation}
is multiplied in both sides of Eq. (\ref{eqn:g1g2}), then
longitudinal and transverse polarizations are considered.
The derivation is too lengthy to be written here, so that the details
should be found in Ref. 4. As a result, we obtain
\begin{align}
g_1 (x, Q^2) = \frac{1}{1+\gamma^2} \int_x^1 \frac{dy}{y}
     \, \big [ \, & \big \{ \Delta f_{1L} (y) + \Delta f_{1T} (y) \big \} 
     \, g_1^{V} (x/y, Q^2)
\nonumber \\
                - & \big \{ \Delta f_{2L} (y) + \Delta f_{2T} (y) \big \} 
     \, g_2^{V} (x/y, Q^2)
  \, \big ]   ,
\label{eqn:g1v}
\end{align}
where the function $\Delta f_i^{VN}(y)$ with $i$=$1L$, $2L$, $1T$,
or $2T$ is defined by 
\begin{equation}
\Delta f_i (y) = f_i^{\lambda_V=+1} (y)
               - f_i^{\lambda_V=-1} (y)  .
\label{eqn:dfy}
\end{equation}
The factor $\gamma^2$ is defined by $\gamma^2 = 4 \, x^2 \, m_N^2 /Q^2$.
The function $f_{1L}^{\lambda_V} (y)$ is the ordinary meson momentum
distribution with the momentum fraction $y$ in the longitudinally
polarized nucleon. There are, however, new contributions from
the $2L$, $1T$, and $2T$ terms. The function $f_{1T}^{\lambda_V} (y)$ is
the distribution in the transversely polarized nucleon.
The functions $f_{2L}^{\lambda_V} (y)$ and $f_{2T}^{\lambda_V} (y)$
are the distributions associated with $g_2$ of the vector meson.
Explicit expressions of $f_i^{\lambda_V}$ are given in
the appendix of Ref. 4.  Our studies are intended 
to investigate a role of the additional terms, 
$2L$, $1T$, and $2T$.

In order to estimate the $g_2^V$ effects,
we approximate it by the Wandzura-Wilczek (WW) relation:
\begin{equation}
g_2^{V (WW)} (x,Q^2) 
             =  - g_1^V(x,Q^2) + \int_x^1 \frac{dy}{y} g_1^V(y,Q^2) .
\end{equation}
Then, the leading-order expression of $g_1^V$ is used for obtaining
\begin{equation}
g_2^{V (WW)} (x,Q^2) =
  \frac{1}{2} \sum_i \, e_i^2 \, [ \, \Delta q_i^{V (WW)} (x,Q^2) 
                          + \Delta \bar q_i^{V (WW)} (x,Q^2) \, ] ,
\label{eqn:ww}
\end{equation}
where the WW distributions are defined by
\begin{equation}
\Delta \bar q_i^{V (WW)}  (x,Q^2) 
            =  -  \Delta \bar q_i^V (x,Q^2) 
               + \int_x^1 \frac{dy}{y} 
                   \,  \Delta \bar q_i^V (y,Q^2) ,
\end{equation}
and the same equation for $\Delta q_i^{V (WW)}  (x,Q^2)$.
In this way, the vector-meson contribution to $\Delta \bar q_i$
in the nucleon becomes
\begin{align}
\Delta \bar q_i^{VNB} (x, Q^2) = \frac{1}{1+\gamma^2} \int_x^1 \frac{dy}{y}
     \, \big [ \, & \big \{ \Delta f_{1L} (y) + \Delta f_{1T} (y) \big \} 
     \, \Delta \bar q_i^V (x/y, Q^2)
\nonumber \\
                -  & \big \{ \Delta f_{2L} (y) + \Delta f_{2T} (y) \big \} 
     \, \Delta \bar q_i^{V (WW)} (x/y, Q^2)
  \, \big ]   .
\label{eqn:dPi}
\end{align}
This equation is used for calculating the $\rho$ meson contributions
to $\Delta \bar u - \Delta \bar d$.
The new terms $2L$, $1T$, and $2T$, are proportional to $\gamma^2$,
namely $1/Q^2$, so that they vanish in the limit $Q^2 \rightarrow \infty$.
Then, Eq. (\ref{eqn:dPi}) agrees on those in Refs. 2 and 3.

%%%%%%%%%%%%%%%%%%%%%%%%%%%%%%%%%%%%%%%%%%%%%%%%%%%%%%%%%%%%%%%%%%%%%%%%%%%%%%
\section{Results}\label{results}

For calculating the obtained expression numerically, the splitting processes
$N \rightarrow \rho N$ and $N \rightarrow \rho \Delta$ are included
with the vertex couplings
\begin{align}
V_{VNN} & =  \widetilde \phi_V^* \cdot \widetilde T \, 
             F_{VNN} (k) \, \,
             \overline u_{N'} \, \epsi^{\mu \, *}
     \,  \bigg[ \, g_V \gamma_\mu 
         - \frac{f_V}{2 m_N} \, i \, \sigma_{\mu\nu} 
                                \khat^\nu  \, \bigg ] \,  u_N ,
\label{eqn:vvnn}
\\
V_{V N \Delta} & = \widetilde \phi_V^* \cdot \widetilde T \, 
             F_{VN \Delta} (k) \, \,
             \overline U_{\Delta,\nu} 
             \, \frac{f_{VN\Delta}}{m_V} \, \gamma_5 \, \gamma_\mu 
             \,  \big [ \, \khat^\mu \, \epsi^{\nu \, *}
             - \khat^\nu \, \epsi^{\mu \, *} \, \big ] \, u_N .
\label{eqn:vvnd}
\end{align}
Here, $\widetilde \phi_V^* \cdot \widetilde T$ indicates the isospin coupling,
$F_{VNN} (k)$ and $F_{VN \Delta} (k)$ are form factors,
$u_N$ is the Dirac spinor, $U_\Delta^\mu$ is the Rarita-Schwinger spinor,
$\epsi^\mu$ is the polarization vector of $\rho$,
$g_V$, $f_V$, and $f_{VN\Delta}$ are coupling constants,
and $\khat^\mu$ is a vertex momentum. 
The momentum $\khat^\mu$ could be taken either 
(A) $(E_V, \vec k)$ or (B) $(E_N - E_B, \vec k)$; however,
the prescription (B) is used in the following results.
For $F_{VNN}$ and $F_{VN \Delta}$, exponential form factors are used
with the 1 GeV cutoff.  
Using these vertices, we calculate the polarized meson momentum
distributions in Eq. (\ref{eqn:dPi}). 

Obtained meson momentum distributions are convoluted with
the polarized distributions in $\rho$. 
The charge symmetry is used for relating the valence quark distributions
in $\rho^-$, $\rho^0$, and $\rho^+$:
\begin{equation}
    (\Delta \bar u)_{\rho^-}^{val}
=   (\Delta \bar d)_{\rho^+}^{val}
= 2 (\Delta \bar u)_{\rho^0}^{val}
= 2 (\Delta \bar d)_{\rho^0}^{val}
= \Delta V_\rho .
\end{equation}
Actual parton distributions are not known in $\rho$, so that they are
assumed as $\Delta V_\rho = 0.6 \, V_\pi$ by considering
a lattice QCD estimate. The distribution in the pion is taken from
the GRS (Gl\"uck, Reya, and Schienbein) parametrization.
Taking into account the isospin factors at the $\rho NN$ and
$\rho N \Delta$ vertices, we obtain
\begin{align}
(\Delta   \bar u   - \Delta \bar d )_{p \rightarrow \rho B}
 = & \bigg [ \, -2 \, \Delta f^{\rho NN}_{1L+1T} 
  + \frac{2}{3} \, \Delta f^{\rho N \Delta}_{1L+1T} \, \bigg ] 
          \otimes \Delta V_\rho
\nonumber \\
- & \bigg [ \, -2 \, \Delta f^{\rho NN}_{2L+2T} 
  + \frac{2}{3} \, \Delta f^{\rho N \Delta}_{2L+2T} \, \bigg ] 
          \otimes \Delta V_\rho^{WW}
\, ,
\label{eqn:ubdb}
\end{align}
where $\otimes$ indicates the convolution integral.

\begin{figure}[h]
\parbox[t]{0.46\textwidth}{
   \begin{center}
%\epsfxsize=0.46\textwidth
%\figurebox{}{}{ubdbn}
\includegraphics[width=0.46\textwidth]{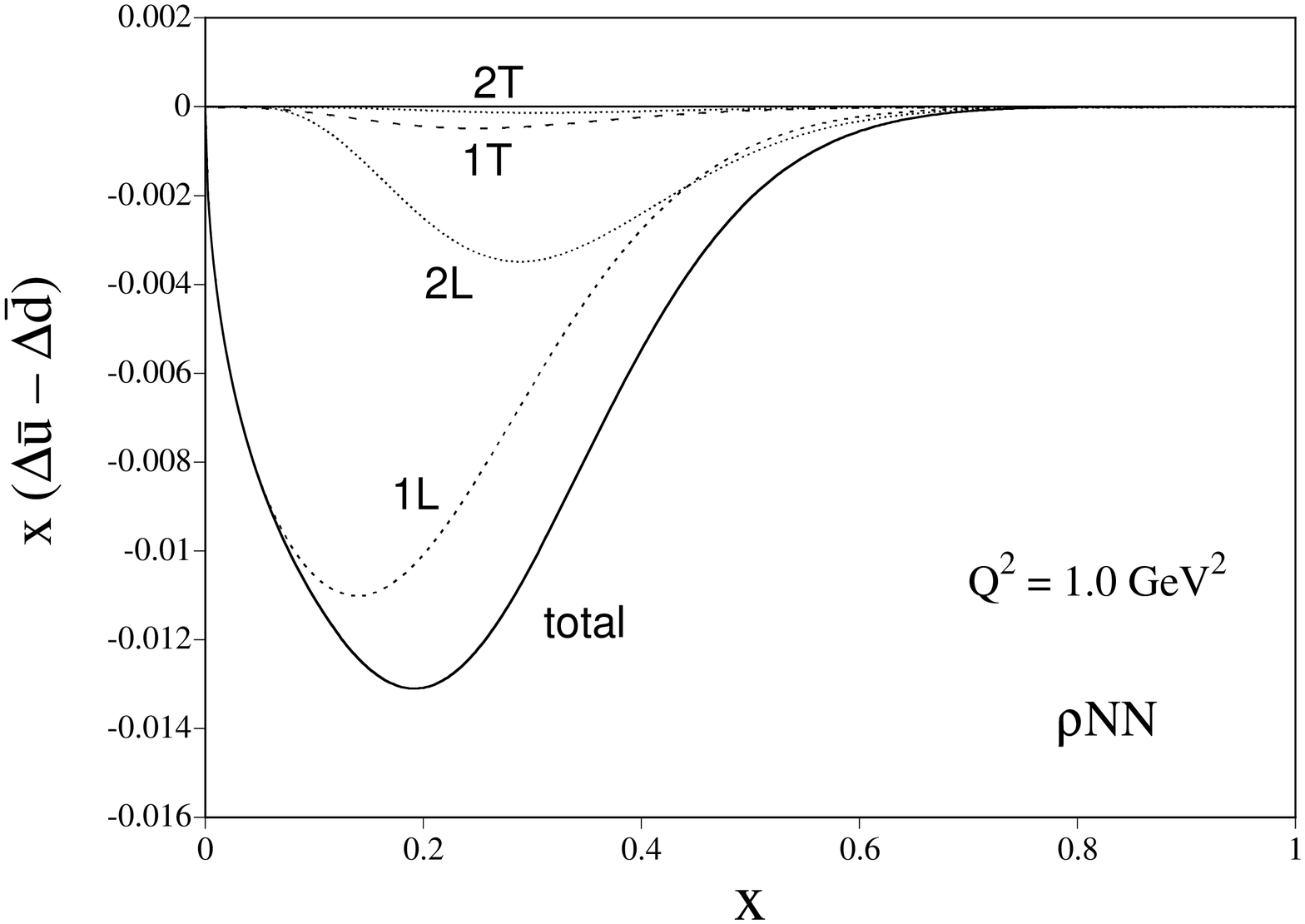}
   \end{center}
   \vspace{-0.7cm}
       \caption{\footnotesize $\Delta \bar u - \Delta \bar d$ from
                              the process $N \rightarrow \rho N$.}
       \label{fig:rhonn}
}\hfill
\parbox[t]{0.46\textwidth}{
   \begin{center}
   \vspace{-0.2cm}
%\epsfxsize=0.46\textwidth
%\figurebox{}{}{ubdbd}
\includegraphics[width=0.46\textwidth]{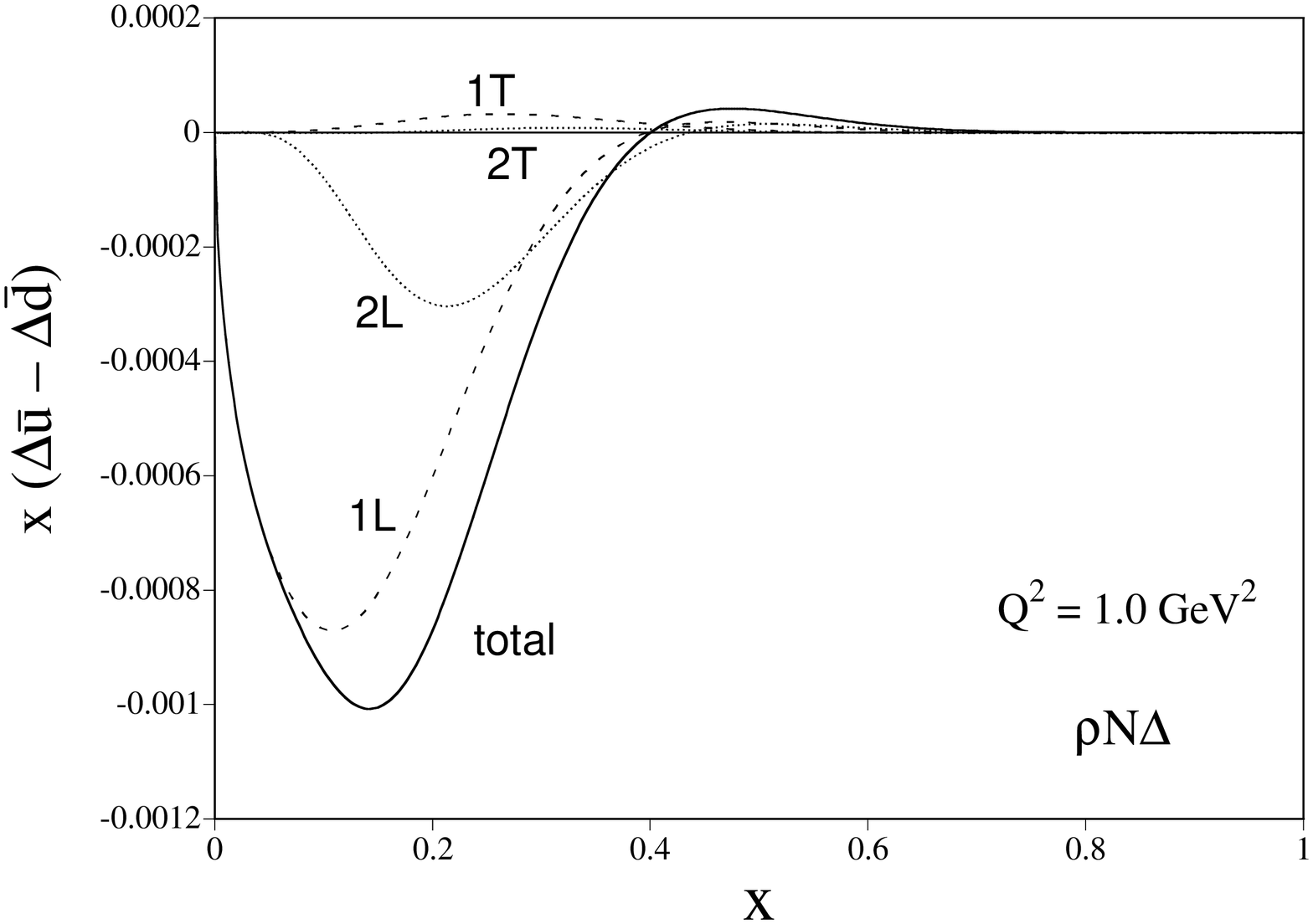}
   \end{center}
   \vspace{-0.7cm}
       \caption{\footnotesize $\Delta \bar u - \Delta \bar d$ from
                          the process $N \rightarrow \rho \Delta$.}
       \label{fig:rhond}
}
\end{figure}

The $\rho NN$ and $\rho N \Delta$ are separately calculated and
numerical results are shown in Figs. \ref{fig:rhonn} and \ref{fig:rhond},
respectively, at $Q^2$=1 GeV$^2$. 
The ordinary longitudinal contributions are denoted as $1L$, and
other new contributions are denoted as $2L$, $1T$, and $2T$.
Among the new terms, $2L$ is the largest one, which becomes
comparable magnitude with $1L$ at medium $x$ ($x>0.2$).
We notice that the distributions from $\rho N \Delta$ are fairly small
in comparison with those from $\rho NN$.
All the distributions in these figures are mainly negative, which
means that $\Delta \bar d$ excess is produced over $\Delta \bar u$
by the meson-cloud mechanism. Because $W$ production and semi-inclusive
processes will be investigated experimentally,
our prediction should be tested in the near future.

\vspace{-0.7cm}
%%%%%%%%%%%%%%%%%%%%%%%%%%%%%%%% figure %%%%%%%%%%%%%%%%%%%%%%%%%%%%%%%%%%%%%%
\noindent
%%%%%%%%%%%%%%%%%%%%%%%%%%%%%%%% figure %%%%%%%%%%%%%%%%%%%%%%%%%%%%%%%%%%%%%%
\vspace{-0.5cm}

%%%%%%%%%%%%%%%%%%%%%%%%%%%%%%%%%%%%%%%%%%%%%%%%%%%%%%%%%%%%%%%%%%%%%%%%%%%%%%
\section{Summary}\label{summary}

The $\rho$ meson contributions to the polarized antiquark flavor
asymmetry $\Delta \bar u - \Delta \bar d$ have been investigated
in a meson-cloud picture. We pointed out especially the existence
of additional terms from the $g_2$ part of $\rho$. The additional
terms become important at medium $x$ with small $Q^2$. 
The obtained $\Delta \bar u - \Delta \bar d$ distributions are
mostly negative, namely the model indicates
$\Delta \bar d$ excess over $\Delta \bar u$, and it should be tested
by future experiments.

%%%%%%%%%%%%%%%%%%%%%%%%%%%%%%%%%%%%%%%%%%%%%%%%%%%%%%%%%%%%%%%%%%%%%%%%%%%%%%%%
%%%%%%%%%%%%%%%%%%%%%%%%%%%%%%%%%%%%%%%%%%%%%%%%%%%%%%%%%%%%%%%%%%%%%%%%%%%%%%%%
\section*{Acknowledgments}
S.K. was supported by the Grant-in-Aid for Scientific
Research from the Japanese Ministry of Education, Culture, Sports,
Science, and Technology. He also thanks his collaborator, M. Miyama,
for discussions.

%%%%%%%%%%%%%%%%%%%%%%%%%%%%%%%%%%%%%%%%%%%%%%%%%%%%%%%%%%%%%%%%%%%%%%%%%%%%%%

%%%%%%%%%%%%%%%%%%%%%%%%%%%%%%%%%%%%%%%%%%%%%%%%%%%%%%%%%%%%%%%%%%%%%%%%%%%%%%
%%%%%%%%%%%%%%%%%%%%%%%%%%%%%%%%%%%%%%%%%%%%%%%%%%%%%%%%%%%%%%%%%%%%%%%%%%%%%%


\begin{thebibliography}{0}
\bibitem{skpr} S. Kumano, Phys. Rep. {\bf 303}, 183 (1998).
\bibitem{fs}  R. J. Fries and A. Sch\"afer,
                   Phys. Lett. {\bf B443}, 40 (1998); hep-ph/9805509 (v3).
\bibitem{cs}  F.-G. Cao and A. I. Signal, 
                   Eur. Phys. J. {\bf C21}, 105 (2001).
\bibitem{km}  S. Kumano and M. Miyama, 
                   hep-ph/0110097 (Phys. Rev. {\bf D65} in press).
\bibitem{cw}  F.-G. Cao, M. Wakamatsu, talks at this conference.
\end{thebibliography}
\end{document}